\RequirePackage{fix-cm}
\documentclass[smallextended]{svjour3}       
\smartqed  
\usepackage{graphicx}
\begin{document}

\title{How to use the Sun-Earth Lagrange points for fundamental physics and navigation
}

\titlerunning{Lagrange Reference Frame}        

\author{A. Tartaglia \and
      E.C. Lorenzini \and \\
      D. Lucchesi \and
      G. Pucacco \and \\
      M.L. Ruggiero \and
      P. Valko
}


\institute{A. Tartaglia \at
              DISAT, Politecnico di Torino and INdAM, Corso Duca degli Abruzzi 24, 10129 Torino, Italy \\
              Tel.: +390110907328\\
              Fax: +390110907399\\
              \email{angelo.tartaglia@polito.it}           
           \and
           E.C. Lorenzini \at
              Department of Industrial Engineering, University of Padova, Via Venezia 1, 35131 Padua, Italy \\
              \email{enrico.lorenzini@unipd.it}
           \and
           D. Lucchesi \at
           Istituto di Astrofisica e Planetologia Spaziali - Istituto Nazionale di Astrofisica (IAPS/INAF), Via Fosso del Cavaliere 100, 00133 Tor Vergata, Roma, Italy \\  and Istituto Nazionale di Fisica Nucleare (INFN), Sezione di Tor Vergata \\
           \email{david.lucchesi@iaps.inaf.it}
           \and
           G. Pucacco \at
           Department of Physics, University of Rome Tor Vergata, Via della Ricerca Scientifica 1, 00133 Rome, Italy \\
           \email{Giuseppe.Pucacco@roma2.infn.it}
           \and
           M.L. Ruggiero \at
           DISAT, Politecnico di Torino and INFN, Corso Duca degli Abruzzi 24, 10129 Torino, Italy \\
           \email{matteo.ruggiero@polito.it}
           \and
           P. Valko \at
           Department of Physics, Slovak University of Technology, Ilkovi\v{c}ova 3, Bratislava 812 19, Slovakia \\
           \email{pavol.valko@stuba.sk}
}

\date{Received: date / Accepted: date}

\maketitle

\begin{abstract}
We illustrate the proposal, nicknamed LAGRANGE, to use spacecraft, located at the Sun-Earth Lagrange points, as a physical reference frame. Performing time of flight measurements of electromagnetic signals traveling on closed paths between the points, we show that it would be possible: a) to refine gravitational time delay knowledge due both to  the Sun and the Earth; b) to detect the gravito-magnetic frame dragging of the Sun, so deducing information about the interior of the star; c) to check the possible existence of a galactic gravitomagnetic field, which would imply a revision of the properties of a dark matter halo; d) to set up a relativistic positioning and navigation system at the scale of the inner solar system. The paper presents estimated values for the relevant quantities and discusses the feasibility of the project analyzing the behavior of the space devices close to the Lagrange points.
\keywords{gravitation \and reference systems \and Sun rotation \and galactic halo}
\end{abstract}

\section{Introduction}
\label{intro}
We propose here to use the Lagrangian ($L$) points of the Sun-Earth system as a physical framework for a number of measurements related to General Relativity (GR) and possible deviations thereof. The same set of $L$ points could furthermore be the basis for a relativistic navigation and positioning system at least at the scale of the inner Solar System.

\begin{figure}[h]
\begin{center}
		\includegraphics[width=8cm]{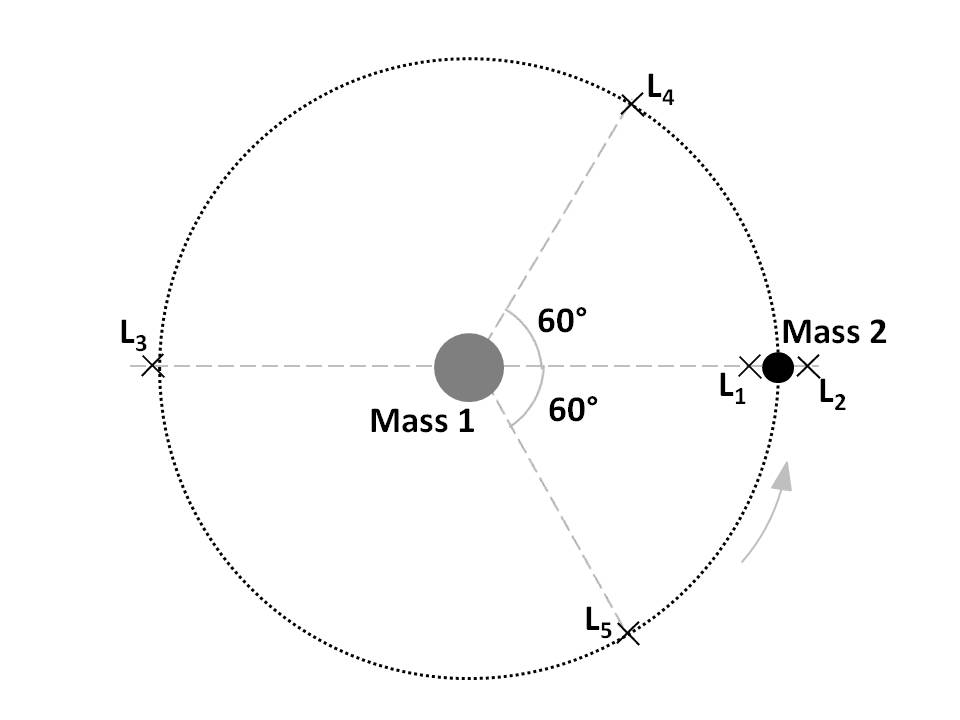}
		\caption{Schematic view of the Lagrangian points of a two body system.}
		\label{fig: 1}
\end{center}
\end{figure}

As it is well known the Lagrangian points of a gravitationally bound two-body system are a
feature of Newtonian gravity. Unlike General Relativity (GR) Newton's
gravity admits analytic solutions for the two-body problem; furthermore,
looking for the positions, on the joint orbital plane, where the attraction
of both bodies on a negligible mass test particle counterbalances exactly
the centrifugal force, one finds five points where such a condition is
fulfilled, with an orbital angular velocity coinciding with that of the two
main bodies around their common center of mass. The traditional labelling of
the five points is $L_1$, $L_2$, $L_3$, $L_4$, $L_5$ and the geometry of the
system is as sketched in Fig. \ref{fig: 1}.

Three points ($L_{1}$, $L_{2}$, $L_{3}$) are saddle points of the effective
potential; in other words, the equilibrium there is unstable, however in the
case of the Sun/Earth pair the instability is very mild. The remaining two
points ($L_{4}$ and $L_{5}$) are real local minima so the equilibrium there
is stable, though corresponding to a shallow potential well. When coming to
a relativistic approach, even though we may guess that the situation is
marginally or even negligibly different from the Newtonian case, the
existence of Lagrangian points is a priori not guaranteed so that the problem
needs a careful discussion. Hopefully the final conclusion is indeed that
libration points (Lagrange points) still exist also in a relativistic 2 plus
1 body configuration, at least in a range of masses including the Sun/Earth
pair, even though no closed solution is available for the position of such
points \cite{Perdomo2016}.

The advantage of the set of the Lagrangian points is that they form a
configuration rigidly rotating together with the Earth. This property has
already been exploited many times for space missions, such as WMAP \cite{Bennett2013}, the
Herschel space observatory \cite{ESA2013}, Planck \cite{Adam2016} (all concluded) and now Gaia \cite{ESA2017}, in $L_{2}$; the Deep Space Climate Observatory \cite{NASA2017}, the Solar and Heliospheric Observatory
(SOHO)\cite{Baldwin2015} and LISA Pathfinder \cite{ESA2015}, in $L_{1}$. The list is not exhaustive and many
more missions are planned directed again to $L_{1}$ or $L_{2}$. It is also worth mentioning that proposals have been issued to exploit, for fundamental physics, the Lagrangian points of the pair Earth-Moon \cite{Battista2015}.

The stability of the positions with respect to one another and to the Earth
makes the Lagrangian points appropriate to work as basis for a physical
reference frame at the scale of the inner solar system. Furthermore,
considering the size of the polygon having the $L$'s as vertices, we may remark
that the time of flight of electromagnetic signals going from one point to
another is in the order of some 10 minutes or more; such long time may act
as a multiplier for the tiny asymmetries originated by angular momentum
effects predicted by GR.

The present paper will discuss the possibilities listed above,
highlighting the advantages for fundamental physics experiments and for the
positioning and guidance of spacecraft out of the terrestrial environment.

In particular, we shall nickname LAGRANGE the proposal of exploiting time of flight measurements along a closed path having $L$ points as vertices, in order to take advantage of the asymmetric propagation produced by the angular momentum of the Sun in the case of two counter-rotating electromagnetic signals. The use of one and the same loop will avoid delays due to different geometric paths for the two beams; the cancellation of the purely geometric component of the time of flight will let the above mentioned asymmetry emerge.

In Section \ref{timedelay} we discuss the GR time delay due both to the Sun and to the Earth; then in Subsection \ref{Delay}, we specialize the analysis to the case where emitter, central body and receiver are aligned (indeed a special case for LAGRANGE). In the calculation, the contribution of the quadrupole of the central body will be included.
In Section \ref{LT} the analysis will concern the more general configuration with a closed contour encompassing a wide area and the purpose will be the measurement of the solar gravitomagnetic effect with an accuracy better than \(1\%\). Section \ref{galactic} evaluates the possibilities to retrieve information about a galactic gravito-magnetic field, if it is there. Section \ref{rps} presents a Relativistic Positioning System (RPS) based on a set of emitters of electromagnetic signals located at the mentioned $L$-points. Since the feasibility of our proposals crucially depends on the possibility of keeping the position of each spacecraft with respect to the corresponding $L$ and its time dependence under control, we shall discuss the issue in Section \ref{Ldyn}. A short conclusion closes the paper.

\section{Gravitational time delay}
\label{timedelay}
A gravitational field produces a time delay and a deflection (deviation) on the propagation of electromagnetic waves.
Presently, we are interested only to describe the effects of the delays in time propagation, because the effects of the deflection on the time propagation are negligible with respect to the leading contributions.
The main effect depends on the mass of the central body and is fully explained in terms of the metric developed by Schwarzschild in 1916 \cite{Schwarzschild1916}. Therefore, these delays are related to the gravitoelectric field of GR \cite{Thorne1988}.
The first successful measurements having the Sun as a source were obtained by Shapiro and collaborators
by means of radar-echoes from Earth to the planets Mercury and Venus \cite{Shapiro1971}. Successively, Anderson and collaborators \cite{Anderson1975} repeated the measurement of the delay in the round trip time from the two spacecraft Mariner 6 and 7 orbiting the Sun.
Finally, Shapiro \cite{Shapiro1977} and Reasenberg \cite{Reasenberg1979} obtained the most accurate results with this technique by means of a transponder placed on the surface of Mars by the NASA mission Viking.
The agreement between the measured delay and its general relativity prediction was around 0.1\%.
 This kind of measurements are quite important because they allow to constrain the PPN parameter \(\gamma\), which measures the space curvature per unit of mass.
Currently, the best measurement of $\gamma$ has been obtained by the radar tracking of the CASSINI spacecraft during a superior conjunction with the Sun along its cruise to Saturn \cite{Bertotti2003}.
Bertotti and collaborators obtained \(\gamma -1\simeq 2\times 10^{-5}\).
The advantage of the latter measurement relies on the Doppler tracking (not exploited
before) and the multi-frequency link (both X-band and Ka-band) that allow for the plasma compensation of the solar corona.
This delay, which is now known as the Shapiro time delay, represents the first GR  correction to the time propagation of an electromagnetic signal between an emitter and a receiver with respect to the time of propagation that is needed in the flat spacetime of Minkowski.

LAGRANGE, with its multi-spacecraft configuration, would allow the measurement of the time delay in the propagation of the electromagnetic signals in several different geometrical configurations.
In the same time, it would extend the measurement of the delay not only to the effect
previously mentioned, the so-called Shapiro time-delay, but also to the delay produced by the gravitomagnetic field \cite{Thorne1988} of the Sun and/or to that of the Earth.
For instance, referring to previous Fig. \ref{fig: 1}, we can consider the propagation of light and the corresponding delay between the two equilateral Lagrangian points \(L_4\) and \(L_5\).
In this case, the impact parameter \(b\), the point of closest approach to the
Sun, is equal to 0.5 AU, i.e. comparable with the other distances, avoiding the problem connected with the plasma of the solar corona, as well as the additional delay produced by the quadrupole moment of the Sun.
Another very interesting geometrical configuration is the one represented by the two
collinear points \(L_1\) and \(L_2\).
The propagation of signals between these two points would allow, for the first time, a direct measurement (in the field of the Earth) of the overall delay on their propagation, as produced by the combined action of the mass and angular momentum of the Earth plus the additional delay due to its oblateness. Some of the corresponding measurements with LAGRANGE would allow to improve the current limits in gravitational physics by exploiting the present know-how and accuracy in time of flight measurements and with the present state of art in atomic clocks precision and accuracy.
Conversely, other effects, in order to emerge from the noise, need an improvement in the current technology of time measurements.
The LAGRANGE measurements would be based on the application of null geodesics around a spinning body in the weak field and slow motion limit (WFSM) of GR.
In terms of metric, the Kerr metric will be the reference \cite{Kerr1963}, or, to say better, its weak field limit \cite{Ohanian2013}, with a non-diagonal component \(g_{0\phi}\) proportional to the intrinsic angular momentum (spin) \(J\) of the central body.

\subsection{Time delays for a configuration where emitter and receiver are aligned with the delaying object}
\label{Delay}
We consider a quasi-Cartesian coordinate system at the post-Newtonian level with origin in the central (deflecting and delaying) body. We consider the propagation in the \(z=0\) plane (coincident with the plane of the ecliptic) and assume that the angular momentum \(\vec{J}\) of the body is along the \textit{z-axis} and that this axis is also the symmetry axis of the body (i.e. we assume cylindrical, or axial, symmetry).
In particular, we assumed a standard isotropic PN approximation \cite{Will1993} where the receiver (or observer) has to be considered positioned along the positive \textit{y-axis}.
Under the above approximations, the line element can be written as:
\begin{eqnarray}
ds^2 = c^2d\tau^2 \simeq g_{00}c^2dt^2 + g_{xx}dx^2 + g_{yy}dy^2 + g_{zz}dz^2 \label{ppn-metric} \\
 + 2g_{0x}dxdt + 2g_{0y}dydt, \nonumber
\end{eqnarray}
where\footnote{Here \(g_{00}\) represents the time-time component of the metric, while the other terms provide spatial and mixed contributions.}
\begin{eqnarray}
g_{00} \simeq -\left(1+2\frac{U}{c^2}\right)\\\label{tempo}
g_{ij} \simeq \left(1-2\frac{U}{c^2}\right)\delta_{ij}\\
g_{0x} \simeq 2\frac{GJ}{c^2r^3}\left(-y\right)\\
g_{0y} \simeq 2\frac{GJ}{c^2r^3}\left(x\right).\label{gravito_yt}
\end{eqnarray}
In the above expressions, \(c\) represents the speed of light, \(\tau\) the (invariant) proper time, \(G\) the Newtonian gravitational constant, \(J\) the angular momentum of the central body, \(r\) the distance in the reference plane, \(\delta_{ij}\) the Kronecker symbol and, finally, \(U\) represents the gravitational potential\footnote{We considered only the main contribution, that arises from the first even zonal harmonic, with respect to the deviation from the spherical symmetry in the mass distribution of the Earth.}
\begin{equation}
U \simeq -\frac{GM}{r}\left(1-J_2\left(\frac{R}{r}\right)^2\frac{3\left(\frac{z}{r}\right)^2 -1}{2}\right),\label{potential}
\end{equation}
where \(M\), \(R\) and \(J_2\) are, respectively, the mass, radius and quadrupole moment of the body.

This configuration is particularly interesting when the delays in the propagation of the electromagnetic signal are analysed in order to take care also of the effect in the time propagation produced by the quadrupole moment of the central object, besides the contributions from the gravitoelectric and gravitomagnetic fields of GR.

In the case of the propagation of electromagnetic waves we need to impose the condition of null geodesics with the further condition that we restrict to the propagation in the reference plane \(z=0\)  with \(x=b\) constant and \(b\ll y\).
Therefore, Eq. (\ref{ppn-metric}) reduces to:
\begin{equation}
0 \simeq g_{00}c^2dt^2 +  g_{yy}dy^2 + 2g_{0y}dydt.\label{ppn-metric-luce}
\end{equation}
We can now solve for the coordinate time element \(dt\) from Eq. (\ref{ppn-metric-luce}) and integrate the final expression from the emitter position at \(y=-y_1\) up to the receiver (or observer) position at \(y=+y_2\) ($y_1$ and $y_2$ are positive quantities and we further assume that $y_2\simeq y_1$).
For the propagation time \(\Delta t_{prop}\) we finally obtain:
\begin{equation}
\Delta t_{prop} \simeq \frac{y_2+y_1}{c} + \frac{2GM}{c^3}\ln\left(\frac{4y_1y_2}{b^2}\right) \pm \frac{4GJ}{c^4b} + \frac{2GM}{c^3}\left(\frac{R}{b}\right)^2J_2 + \dots,\label{delta_t}
\end{equation}
where smaller contributions to the time delay have been neglected.

The first term in Eq. (\ref{delta_t}) accounts for the time propagation in the flat spacetime of Special Relativity. The second term represents the contribution from the gravitoelectric field of GR in the weak field approximation: it is the Shapiro time delay. The third contribution arises from the gravitomagnetic field in the same approximation. The \(\pm\) sign accounts for the chirality of this contribution: it is positive for a propagation of the signals in the same sense of rotation of the central mass, it is negative in the case of the opposite sense for the propagation.
Finally, the last term represents the contribution that arises from the oblateness of the central body.

The solution provided in Eq. (\ref{delta_t}) implies that emitter, central body and receiver have the same \(x\) and \(z\) coordinates (with \(x=b\), \(z=0\)) and differ only for the \(y\) coordinate (which is negative for the emitter, null for the central body and positive for the receiver).

The result obtained in Eq. (\ref{delta_t}) can be considered as a particular case of two results obtained in previous works \cite{Ciufolini2002,Ruggiero2002}.
In fact, our result coincides with that obtained in \cite{Ciufolini2002} when that work is restricted to the lensing effect which can be obtained for light propagating in their symmetry plane (coincident with our reference plane) with the transformations \(\gamma=0\) and \(\beta=\pi/2\) in their expressions and with the further condition \(\alpha=0\) or \(\alpha=\pi\) in their final expressions for the delays due to the angular momentum and the quadrupole coefficient (see in particular their section 2).\footnote{Here \(\gamma\) (not to be confused with the PPN parameter commonly designated by the same symbol) and \(\beta\) represent two of the Euler angles that define the orientation of their symmetry plane with respect to the lens plane, while \(\alpha\) represents the angular position of a generic light ray over the lens plane. }
Conversely our first three terms in Eq. (\ref{delta_t}) coincide with equations (55), (56) and (57) of \cite{Ruggiero2002} with the transformation \(y_1 \rightarrow -y_1\) for their \(y_1\) and the approximation \(y_2 \gg b\) and \(y_1 \gg b\) for their coordinates.

By applying the measurements based on the propagation time determined with Eq. (\ref{delta_t}) to the configuration \(L_1\)--Earth--\(L_2\), it will be possible (at least in principle) to obtain a measurement of the Earth's quadrupole coefficient in a way independent from the usual space geodesy techniques based on the inter-satellite tracking --- by means of the two twin GRACE satellites \cite{Reigber2005} --- and from the precise orbit determination (POD) of laser-ranged satellites in orbit at a relatively high altitude, as in the case of the two LAGEOS \cite{Cheng2013}.
Considering that the distance between the two Lagrangian points $L_1$ and $L_2$ is $y_1+y_2\simeq 3\times10^9$ m, and assuming an impact parameter \(b\) of the order of the Earth's radius \(R_{\oplus}\simeq 6.4\times 10^6\) m, for the propagation time of Eq. (\ref{delta_t}) we obtain:
\begin{equation}
\Delta t_{prop} \simeq \left( 10 s\right) + \left(3.6\times10^{-10}s\right) +\left(\pm 3\times 10^{-17}s\right) +\left(3.2\times 10^{-14}s\right)+ \dots,\label{delta_t-L1L2}
\end{equation}
where the contribution of each term has been highlighted.
If we consider a round trip travel for the propagation time, the smaller contribution of the gravitomagnetic field cancels out when we consider the propagation on the same side of the Earth, and the quadrupole effect can be extracted after modelling the Shapiro delay and the larger effect of the propagation time in the flat spacetime of Minkowski.

The knowledge of the oblateness of the Earth is particularly important because of its long-term variations in relation to the Earth's internal structure and its mass distribution. In fact, phenomena like the melting of glaciers and ice sheets as well as mass changes in the oceans and in the atmosphere are responsible for variations in the rate of the global mass redistribution with a consequent time dependency in the quadrupole coefficient characterized by annual and interannual variations \cite{Cheng2013}.

This kind of measurement can be initiated by Earth, the delaying body, with all the advantages of an Earth based Laboratory equipped with the best time-measuring apparatus to perform the experiment.
In particular, optical clocks and lattice clocks based on \(Sr\)-atoms have reached  outstanding fractional frequency instabilities down at a level of about \(2\times 10^{-16}/\sqrt{T}\) or less, with \(T\) the integration time \cite{Al-Masoudi2015,Nicholson2015}.

For instance, with an integration time of about \(10^4\) s it is possible to reach a precision in the measurement of the quadrupole coefficient of about \(\delta J_2/J_2\simeq 3\times 10^{-8}\), comparable with the current best determinations from Earth (with calibrated errors) using the LAGEOS' data, and even better with longer integration times. Considering that the time of flight between $L_1$ and $L_2$ is of the order of $10$ $s$, longer integration times imply of course a number of bounces back and forth between the end points of the trajectory.

\section{Solar Lense-Thirring drag}
\label{LT}
The Lense-Thirring effect (LT) or inertial frame dragging by a moving massive body is a feeble effect of GR, first considered by Thirring \cite{Thirring1918}  and Lense and Thirring \cite{Lense1918} in 1918, while studying the influence of rotating masses (in particular a rotating hollow massive spherical shell) on a test particle. LT may also be considered as a manifestation of gravito-magnetism i.e. of that typical component of the GR gravitational interaction resembling the magnetic field of moving charges.

So far, LT has been verified experimentally in a limited number of cases. A careful analysis of the orbits of the LAGEOS and LAGEOS 2 satellites, monitored by laser ranging, evidenced the LT drag of the nodes of the orbits with a $10\%$ accuracy \cite{Ciufolini2004}\cite{Ciufolini2011}.The Gravity Probe B experiment measured the precession induced by the gravitomagnetic field of the Earth on four orbiting gyroscopes, with a $19\%$ accuracy \cite{Everitt2011}. The ongoing LARES experiment (combined with the previous data from the two LAGEOS) has attained a preliminary $5\%$ accuracy \cite{Ciufolini2016}.
With a different technology, the GINGER experiment is under study and preliminary test of the technology. It is based on the use of an array of ring lasers to be located underground at the National Gran Sasso Laboratories in Italy \cite{Bosi2011,Divirgilio2014,Tartaglia2017}. Ring lasers are extremely sensitive rotation measuring devices. Their operating principle is a GR evolution of the old Sagnac effect \cite{Sagnac1913}; what is measured in practice are frequency and amplitude of a beat between two stationary counter-propagating light beams in the ring. Rotations, either of kinematical origin or due to the chirality of the gravitational field (gravitomagnetic component), produce a right/left asymmetry of the propagation along the ring. The aim of GINGER is to verify the terrestrial LT within $1\%$ or better.

The use of the Sun-Earth Lagrangian frame would allow a measurement of the solar gravitomagnetic field (solar LT), exploiting the Sagnac approach but resorting to time of flight measurements rather than to interference phenomena or beat tones. For our purpose we may start from the external line element of a steadily rotating body in a reference frame where the main mass is at rest and the axes do not rotate with respect to the distant stars (to the quasars). As in the previous section, weak field conditions are assumed, but now, for convenience, we use polar coordinates in space. It is:

\begin{eqnarray}
ds^2=(1-2\frac{m}{r}) c^2 dt_0^2-\frac{dr^2}{(1-2\frac{m}{r})}-r^2 d\theta^2-r^2 (sin{\theta})^2 d\phi_0^2 \nonumber \\
+4\frac{j}{r} sin^2{\theta}c dt_0 d\phi_0
\label{line0}
\end{eqnarray}

The quadrupole moment of the main body has been neglected. If $M$ is the mass of the source, it is $m=GM/c^2$ with the dimension of a length. Similarly, if $J$ is the modulus of the angular momentum of the source, it is $j=GJ/c^3$ with the dimension of a squared length. The index $0$ labels the coordinates specific of the non-rotating, asymptotically flat reference frame.
In the case of the Sun $m_{\odot} = 1475$ m and $j_{\odot} = 4.7144\times 10^6$ m$^2$.

It is convenient to use coordinates apt for a terrestrial (or co-orbiting with the Earth) observer. In practice we need to combine a rotation of the axes at a rate $\Omega$ corresponding to the orbital motion of the Earth, together with a boost at the tangential speed of the Earth on its orbit $V$ \cite{Divirgilio2010}. What holds for the Earth, holds for the Lagrangian points too. Since we are considering free fall the orbital rotation rate is Keplerian, so that:

\begin{equation}
\hspace{1cm} \Omega=c\sqrt{\frac{m}{a^3}};\hspace{2cm} V=\Omega a=c\sqrt{\frac{m}{a}}
\label{kepler}
\end{equation}

Here $a$ is the radius of the orbit of the Earth ($\sim 1.5\times 10^{11}$ m), $m_\odot /a \sim 10^{-8}$, and $j_\odot/a^2 \sim 10^{-16}$.

We may now restrict our attention to the orbital plane, so that it is $\theta=\pi/2$. In the new reference frame and with the new coordinates (see the Appendix for the details) the line element is

\begin{eqnarray}
ds^{2} &\simeq& \left[1+\frac{m}{a}\left(1+\frac{m}{a}\right)\left(1-2\frac{a}{r}\right)\right]c dt^2 \nonumber \\
&&-
\left(1+2\frac{m}{r}+4\frac{m^2}{r^2}\right)dr^2 \nonumber\\
&&-
\left[1-a\frac{m}{r^2}-\frac{m^2}{r^2}\left(1-2\frac{a}{r}\right)\right]r^2d\phi^2 \label{line1} \\
&&+
2\left[2\frac{j}{ra}-\sqrt{\frac{m}{a}}-\left(1-2\frac{a}{r}\right)\left(\frac{m}{a}\right)^{3/2}\right]a c dt d\phi \nonumber
\end{eqnarray}

The approximation has been kept to the lowest order in $j$ and with reference to the numerical values holding for the Sun. For short we write:
\begin{equation}
g_{0\phi }=c \left[2\frac{j}{ra}-\sqrt{\frac{m}{a}}-\left(1-2\frac{a}{r}\right)\left(\frac{m}{a}\right)^{3/2}\right]a
\end{equation}

The frame is non-inertial and comoving with the laboratory; the origin
remains in the center of the Sun.\footnote{It should actually be in the barycenter of the Sun-Earth pair, but the difference should be discussed among the perturbations of the spherically symmetric system.}

In order to find out the time of flight of an electromagnetic signal along a given path, we may extract the time element from Eq. (\ref{line1}) remembering that it is $ds=0$. In terms of a generic stationary axially symmetric space-time and referring to general coordinates, we find:

\begin{equation}
c dt=\frac{-g_{0i}dx^i\pm \sqrt{(g_{0i}dx^i)^2-g_{00}g_{ij}dx^i dx^j}}{g_{00}}
\label{ds0}
\end{equation}

In order to insure an evolution towards increasing real times, the $+$ sign must be chosen. Then we see that the term containing the square root in the right hand side of Eq. (\ref{ds0}) does not change sign when reversing the sense of motion along a given path, whereas the first term in the numerator does. Since we are interested in the asymmetries in the propagation we consider the difference between the right- and left-handed time of flight along the same elementary section of the path; in this way the square root cancels and the other term doubles. Finally we integrate along the whole closed space trajectory and express the result in terms of the proper time of the observer. The total time of flight asymmetry turns out to be \cite{Ruggiero2015}:

\begin{equation}
c \delta \tau=-2\sqrt{g_{00}} \oint{\frac{g_{0i}}{g_{00}}dx^i}
\label{tofd}
\end{equation}

\subsection{Application to a Lagrangian polygon}
\label{subsec:polygon}

Casting into Eq. (\ref{tofd}) the information extracted from Eq. (\ref{line1}) and preserving the solar weak field approximation, we get:

\begin{eqnarray}
c\delta \tau &\simeq& -2\left\{1+\frac{m}{2a}\left(1-2\frac{a}{r}\right)\left[1+\frac{m}{2a}\left(\frac{3}{2}+\frac{a}{r}\right)\right]\right\}\allowbreak
\times \nonumber \\
& & \oint\left(2\frac{j}{r^2}\right)rd\phi  \label{tofu} \\
&\simeq& -4\oint\frac{j}{r}d\phi \nonumber
\end{eqnarray}

Suppose now that the closed path is a polygon, whose edges are light rays. Of course the corresponding null trajectories will be affected by the gravitational lensing due to the mass of the Sun. However we know that the angular deviation due to the lensing effect is proportional to $m_\odot$, so that its influence in the calculation of (\ref{tofu}) is negligible. In practice we may assume the space trajectories of electromagnetic signals to be straight; the typical equation is simple:
\begin{equation}
\frac{b}{r}=\cos{(\phi-\Phi)}
\label{straight}
\end{equation}
The closest distance from the straight line to the center of the system is $b$ and the azimuth of the closest point is $\Phi$.

Suppose for example that a signal, propagating in the ecliptic plane, goes from position $A$, with coordinates $r_A$ and $\phi_A$ to the arrival point $B$ with coordinates $r_B$ and $\phi_B$. We easily work out the contribution of this stretch to the integral (\ref{tofu}):

\begin{equation}
c \delta \tau_{AB}\simeq 4\frac{j}{b}\left(\sin{(\phi_B-\Phi)}-\sin{(\phi_A-\Phi)}\right)
\label{lato}
\end{equation}

Let us apply the above result to a triangular loop having $L_4$, $L_2$ and $L_5$ at the corners. The coordinates in the plane of the ecliptic, measuring the angles from the Sun-Earth line, are:
\begin{eqnarray}
L_4&:&\hspace{1 cm} r_4=a, \hspace{2 cm} \phi_4=\pi/3 \nonumber \\
L_2&:&\hspace{1 cm} r_2=a+a_2,\hspace{1.5 cm} \phi_2=0 \nonumber  \\
L_5&:&\hspace{1 cm} r_5=a, \hspace{2 cm} \phi_5=-\pi/3 \nonumber
\label{lati}
\end{eqnarray}

It is $a_2\sim 1.5\times 10^9$ m, so that $a_2/a\sim 10^{-2}$. The minimum distance between the $L_4-L_2$ (or $L_5-L_2$) line and the center of the system is
 \begin{equation}
 b_{24}=b_{25}=(a+a_2)\cos{(\frac{\pi}{6}+\frac{\sqrt{3}}{2}\frac{a_2}{a})}
 \label{bi}
 \end{equation}

 Numerically: $b_{24}=b_{25} \sim 1.3\times 10^{11}$ m. The angular coordinate of the minimum distance point, on one side or the other, is $\Phi_{24}=-\Phi_{25}\simeq \frac{\pi}{6}+\frac{\sqrt{3}}{2}\frac{a_2}{a}$.

 Coming to the $L_4-L_5$ line, it is
 \begin{equation}
 b_{45}=a\cos{\frac{\pi}{6}}\simeq 7.5\times 10^{10} \textsf{ m}
 \label{45}
 \end{equation}
  and $\Phi_{45}=0$.

Summing up, and considering the full triangle, we have:
\begin{eqnarray}
\delta \tau_{245}&=& 2\delta \tau_{52}+\delta \tau_{45} \nonumber \\
&\simeq& 8\frac{j\sqrt{3}\sin{\frac{\sqrt{3}}{2}\frac{a_2}{a}}}{c (a+a_2)\cos{(\frac{\pi}{6}+\frac{\sqrt{3}}{2}\frac{a_2}{a})}} \allowbreak
-8\frac{j}{\sqrt{3}c a} \nonumber \\
\label{triangolo}
\end{eqnarray}

Eq. (\ref{triangolo}) may be approximated to first order in $a_2/a$ i.e. at the $\%$ level:
\begin{equation}
\delta\tau\simeq 8\sqrt{3}\frac{ja_2}{c a^2}-8\frac{j}{\sqrt{3}c a}
\label{triapp}
\end{equation}

Finally, casting numbers in, we obtain (in seconds):
\begin{equation}
\delta \tau_{245}\simeq 4.30\times 10^{-13}
\label{number}
\end{equation}

The total expected time of flight asymmetry is well within the range of measurability, at least in terrestrial laboratory conditions. The challenge is to measure it in space.

\subsection{Retrievable information on the interior of the Sun}
\label{interior}
Besides making use of Sun's angular momentum as a source of a LT field for a basic science experiment, i.e. for another precise test of GR, there are other, some even truly practical, reasons for such observations. It is widely believed that the observed differential rotation of the Sun triggers a near-surface layer of rotational shear, known as \textit{tachocline}, where large-scale dipole magnetic fields are generated by dynamo action, ultimately leading to the 11-year solar cycle of sunspots \cite{Hughes2012}. Crucial to the possible role of the tachocline in the dynamo are its location and depth. While Sun's photosphere can be directly observed and also neutrinos provide some direct information about processes in the core of our star, the tachocline is not directly observable. Until now, all available information about this boundary layer between the radiative interior and the differentially rotating outer convective zone, have been collected via helioseismology observations, mainly using the Solar and Heliospheric Orbiter (SOHO) and the Solar Dynamics Observatory (SDO) probes \cite{SOHO2004,Christensen2002}. The estimated location of the shear layer at Sun's equator is ($0.693\pm0.002) R_\odot$, i.e. beneath the convection zone base, and with a width of $0.04 R_\odot$. Using Sun's density profile based on the Standard Solar Model \cite{Bahcall1989,Turck2016}, the tachocline itself should contribute at the level $\sim 0.5 \%$ to the total angular momentum of the Sun, i.e. to the source of the LT field. Although such precision of the solar LT field determination lies at the very edge of the expected LAGRANGE project sensitivity, a periodic low frequency temporal variation of the LT field strength would open another window for Sun interior studies.

\section{Relevance of the measurement of a possible galactic gravitomagnetic field}
\label{galactic}
When addressing the effects of rotating massive bodies (Sun) on a local space-time geometry in our planetary system, it is rational to consider also possible analogous effects originating from larger structures dynamics, i.e. from our Galaxy or even more.
The main reason is that fields associated with metric tensor components ($g_{0i}$ or $g_{0\phi}$ as used in Eq.s (\ref{ppn-metric}) and (\ref{line0})) might mimic the effects typically associated to the presence of dark matter (DM), i.e. additional centripetal or centrifugal acceleration ($a_c \propto v B_{LT} $) and gravitational lensing (effective refraction index $n \propto 1-A_{LT}$). The gravito-magnetic field potential ($A_{LT}$) and field strength ($B_{LT}$), rather than metric tensor components, are used for clearer analogy only.
The best studied and quantified DM problem is related to the dynamic stability of dwarf and spiral galaxies. In the case of the Milky Way (MW) the distribution of the accounted for luminous mass in stars ($\sim 5\times 10^{10} M_\odot$), nonluminous interstellar gas and dust ($\sim 5\times 10^9 M_\odot$), central black hole  ($\sim 4\times10^6 M_\odot$)\footnote{In the center of our galaxy, there is an extremely dense compact object (Sagittarius A*) most probably consisting of a black hole.} and central bulge ($\sim4.5\times10^9 M_\odot$) is not compatible with the observed nearly flat rotation curves [$v(r)=const$] of stars and gas in the disk \cite{Binney2008}.
The same property of rotation curves for spiral galaxies has been confirmed for star-free, edge regions, via radio emission observations of neutral hydrogen \cite{Sofue2001}.
For the MW, the mutual gravitational attraction of stars, central black hole and interstellar dust provide a significant part (nearly all) of the required centripetal force at small distances (up to $5$ kpc) from the center, while flat rotation curves at larger distances (above $10$ kpc) undeniably point towards some other source of centripetal force \cite{Iocco2015}.

The typical approach to address this problem is to postulate the presence of a large galactic halo, extending beyond $30$ kpc, consisting of massive nonluminous particles with isothermal spherical distribution. In such models, the stabilizing effect of DM is being contemplated through a static gravitational field (i.e. the $g_{00}$ component of the metric tensor) due to the mass of the invisible halo. Depending on the peculiar DM model, the total mass of the MW may be entirely dominated by the dark halo and could reach values ranging from $\sim5.2\times10^{11} M_\odot$ \cite{Eadie2016} up to $\sim1.5\times10^{12} M_\odot$ \cite{McMillan2011}.

Using a naive but straightforward example, a LT field of $\sim 8.9\times10^{ - 16 }$ s$^{ - 1}$ strength, would suffice to provide all necessary centripetal acceleration to account for the motion of the Sun around the MW center with a tangential velocity $220$ km/s at $8$ kpc radius. According to a realistic MW mass distribution model \cite{Iocco2015,Gerhard2002}, an additional force component is needed, to account for $30$ km/s of the total $v_{LSR} = 220$ km/s orbital velocity.\footnote{LSR stands for \textit{Local Standard of Rest}.} A local value of the LT field $B_{LT}\sim 2.2 \times 10^{ - 16 }$ s$^{ - 1 }$ would account for this additional centripetal force. Similar values could be deduced for the M31 galaxy \cite{Carignan2006} in which the rotational velocity term associated with the DM scenario provides a linearly growing contribution to the rotation curve with a slope of $1.2 \times 10^{ - 16 }$ s$^{ - 1}$. This result could also be interpreted as the influence of a homogenous LT field, perpendicular to the plane of the disk of the galaxy, with an identical intensity of $B_{LT}\sim 1.2 \times 10^{ - 16 }$ s$^{ - 1}$. The hypothesized LT field strengths are weaker than the current experimental possibilities \cite{Everitt2011,Ciufolini2011} but well within the LAGRANGE project scope. On the contrary, the galactic LT field strengths, calculated from known baryonic mass-velocity data, are typically much weaker: only $\sim 2.6 \times 10^{ - 22 }$ s$^{ - 1}$ for the MW at the Sun distance from the center.
Taking into account that none of the recent experiments have been able to detect the physical nature of DM (see overview in \cite{Strigari2013}) and a clear observational evidence of a strong correlation between galactic baryonic content and plateau velocity of the rotation curves ($v_p$), expressed through the Tully-Fisher relation ($M_B \propto v_p^4$) \cite{Tully1977} for more than $100$ rotationally supported galaxies of different masses and morphologies \cite{McGaugh2016}, constitutes a justified reason to consider alternatives to standard DM scenarios and assess the possible presence of local LT fields with strength in the $10^{ - 16 } \div 10^{ - 20 }$ s$^{ - 1}$ range.

Overwhelming share of non-baryonic energy-mass density, in coincidence with observed clustering of visible matter over a larger volume in the observable universe, suggests that other potential sources of LT fields, with strengths within the interval of interest, are viable. In that respect, even a residual primordial LT field, originating from the initial singularity and the subsequent fast evolution processes (inflation era) cannot be excluded. Presumed primordial LT fields would imprint on the CMB spectrum in a similar way as DM and would influence large scale structures evolution into characteristic filaments with congregated clusters of galaxies and large voids in between \cite{Tully2014}. Similar filaments and voids structures and other relevant analogies are commonly observed in solid state systems \cite{Volovik2003}. Filaments of vortex lines of quantum vortices in superfluid helium or magnetic flux bundles in superconducting materials are the best examples. Although such simple analogies cannot guarantee they would have something in common with large scale structures in the universe, the existence of forces inside and among such filaments, originating from the interaction with the bulk of the medium (space-time voids), looks a lot like DM and dark energy (DE) effects. Therefore local LT field search (measurement) might open another window into the DM and DE problem.

\section{Relativistic positioning}
\label{rps}
The solution adopted for global positioning on Earth or in its vicinity is
mainly based on the GPS method and on the GPS, GLONASS, Galileo and other
present or future satellites dedicated constellations. Without entering into
a discussion of the strengths and weaknesses of that approach, it is easily
agreed that it cannot be extended beyond the near terrestrial environment
or, at least, that the application to space navigation is an opportunity to
reconsider the whole method, especially regarding the way to account for the
effects of special and general relativity.

\begin{figure}
	\begin{center}
		\includegraphics[width=8cm]{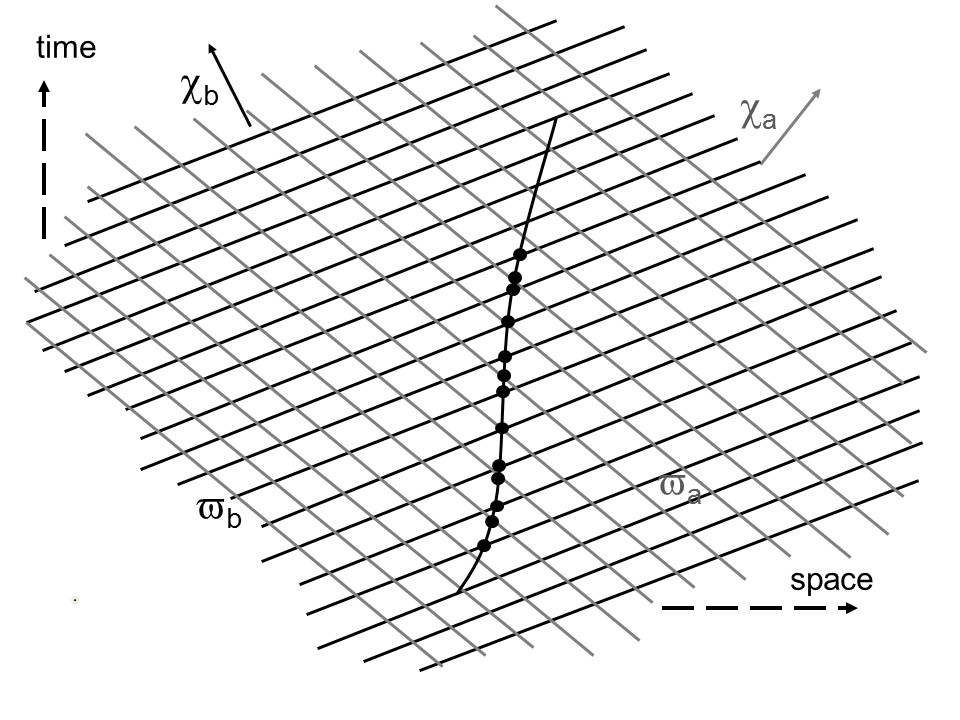}\\
		\caption{Bidimensional example of the positioning method outlined in the text. The $\chi$'s are the null wavevectors of the signals coming from the two (four in the full space-time) independent sources. The $\varpi$'s label the null wavefronts of the pulses. The wiggling line is the worldline of the receiver. The dots show some of the intersection points (reception of a signal). The relevant quantity is the length (i.e. proper time span) between successive reception events.}
		\label{Fig. 2}
	\end{center}
\end{figure}

An intrinsically relativistic positioning system (RPS) has been proposed and
is described in \cite{Tartaglia2013,Tartaglia2011}. It is based on the local timing of at least four remote independent sources of electromagnetic pulses; the essence of the method is graphically presented in Fig. \ref{Fig. 2}. Successive pulses (but they could also be periodic equal phase hypersurfaces) cover space-time by a regular four-dimensional lattice. The world-line of an observer intersects the walls of successive cells of the lattice; the proper time interval measured by the  observer between consecutive crossings provides the basic information. Counting the pulses (after identifying the various sources) and applying a simple linear algorithm it is possible to calculate the coordinates (including time) of the receiver in the fiducial reference frame \cite{Tartaglia2011}.

The dimensionless coordinates along the light cone of a source are expressed as the sum of an integer part $n_a$ (the subscript $a$ labels the sources) and of a fractional part $X_a$. The integer is obtained just counting the successive arrivals of the pulses; the fractional part is given by a simple linear algorithm applied to sequences of arrival times in the proper reference of the observer \cite{Tartaglia2011}. Projecting the $n_a+X_a$ light cone coordinates onto the axes of the fiducial reference frame finally produces the practical coordinates we are interested in.

Of course the sources may be orbiting satellites, but in that case you have to know with the best
possible accuracy, the position of each satellite (i.e. its real orbit)
while time passes. The situation would be far simpler if the position of the
emitter were fixed in the fiducial reference frame. This possibility is implemented in nature if the signals come, for
instance, from pulsars: their positions in the sky are indeed fixed or
slowly moving at a well known rate; furthermore pulsars are also very good
clocks, even better, in the long term, than our atomic clocks. An exercise
application of the RPS, using pulsars, is presented in \cite{Ruggiero2011}.

The inconvenience of pulsars is that their pulses are extremely weak so that
large antennas are needed and special techniques must be implemented in
order to identify and extract the signal from an overwhelming noise. Such
troubles can be removed placing artificial "pulsars" in points that keep
rigidly their positions in an appropriate reference system. That is indeed
the case of the Lagrangian points. $L_1$, $L_2$, $L_4$ and $L_5$, equipped with emitters of regular pulses would form a very interesting basis for a physical reference frame co-orbiting with the Earth. $L_3$ has not been considered because it is located in the opposite side of the Sun with respect to the Earth, so being invisible from our planet. An important feature of the system is that the distances between the reference points range between $1.5$ million km approximately (from $L_1$ or $L_2$ to the Earth) to $150$ million km (from the Earth to $L_4$ or $L_5$). Such large values dramatically reduce the effect of the geometric dilution which renders GPS (and the other terrestrial positioning systems) useless when extended away from our planet. Of course all Lagrangian points lie in a plane and that is usually the case also for most space missions, but the size of the base and using four $L$ points reduces the problem of geometric dilution within distances of a few AU, excepting limited "wakes" along the lines containing a couple of emitters.

The spacecraft carrying the emitter devices could in general not coincide with the corresponding Lagrange point, but would rather orbit around the point on stable ($L_4$ and $L_5$) or on halo or weakly unstable Lissajous orbits ($L_1$ and $L_2$). The final accuracy of the positioning would depend mainly on the accuracy with which the instantaneous position on the orbit is known; we are discussing this issue in the next section.

The other limiting factor for the final result is the quality of the clock used by the receiver: in principle a clock fit for a $10^{ - 10}$s accuracy attains also a centimeter accuracy in determining a travelled distance.

\section{Orbital dynamics around the $L$ points}
\label{Ldyn}
The transmitting/transponding spacecraft of LAGRANGE will be placed in orbits around the collinear and triangular Lagrangian points.  The motion around the Lagrangian points of a small body is described by the classical solution of the Restricted Three Body Problem (RTBP) \cite{Szebehely1967}.  The assumptions that underline the RTBP are that the orbiting body has a negligible mass with respect to the two primaries, in our case Earth and Sun, and the primaries follow circular orbits. We further restrict our preliminary assessments to the planar case of the RTBP.

It is well known that motion around a collinear point is always (weakly) unstable while the stability of motion about the triangular points depends on the mass ratio of the two primaries.  In the case of the Sun-Earth system (and any other combination of mass ratios in the solar system) the motion is stable.  LAGRANGE will require spacecraft placed in orbit around $L_1$, $L_2$, $L_4$ and $L_5$.  The planar motion in the proximity of the Sun-Earth $L_4$ and $L_5$ (where a linear approximation holds) follow orbits that have two frequency components: a faster motion with a 1-year period and a much slower one with a 156-year period.

\begin{figure}
\begin{center}
		\includegraphics[width=8cm]{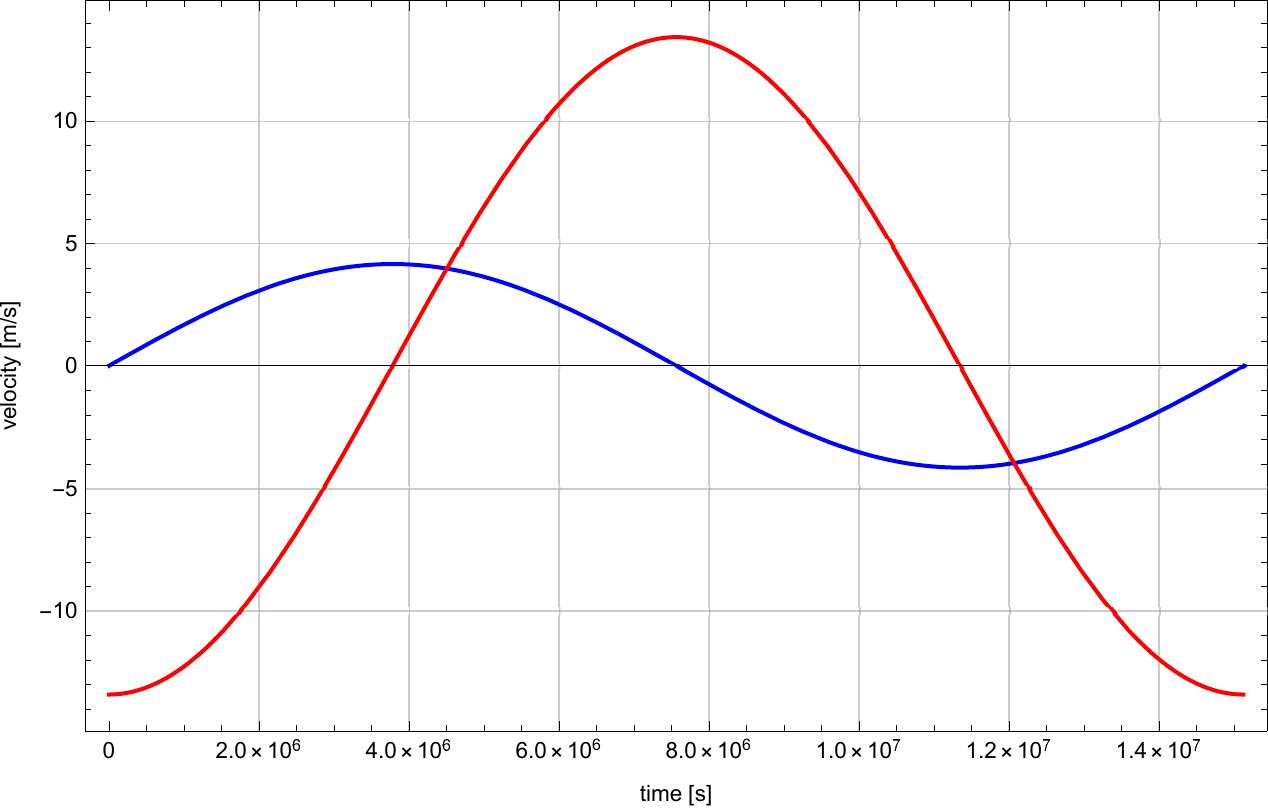}
		\caption{Velocity components along the synodic axes of a 10-km wide
orbit around $L_1$. Very similar figures are obtained for an orbit with the same amplitude at $L_2$.}
		\label{Fig. uno}
\end{center}
\end{figure}

The planar motion around a collinear point is characterized by a couple of complex-conjugate eigenvalues associated with a "stable" manifold and a couple of positive real roots associated with an unstable manifold that produces a divergent motion of the spacecraft.  Initial conditions can be chosen in such a way as to excite only the complex conjugate eigenvalues in order to minimize the instability that however will eventually be excited by non-gravitational perturbations or earlier on by imperfect initial conditions.  For the Sun-Earth system, the period of the "weakly-unstable" planar orbit is half a year. Including also the out-of-plane component one gets a second but very close period.  The different frequencies give rise to orbits that describe quasi-periodic Lissajous figures or, for sufficiently large amplitude, inclined "halo" orbits. Fig. \ref {Fig. uno} shows the velocity components of a 10-km wide orbit in the stable manifold of $L_1$. Due to the small amplitude, an orbit with the same amplitude around $L_2$ has very similar velocity values.

\begin{figure}
\begin{center}
		\includegraphics[width=8cm]{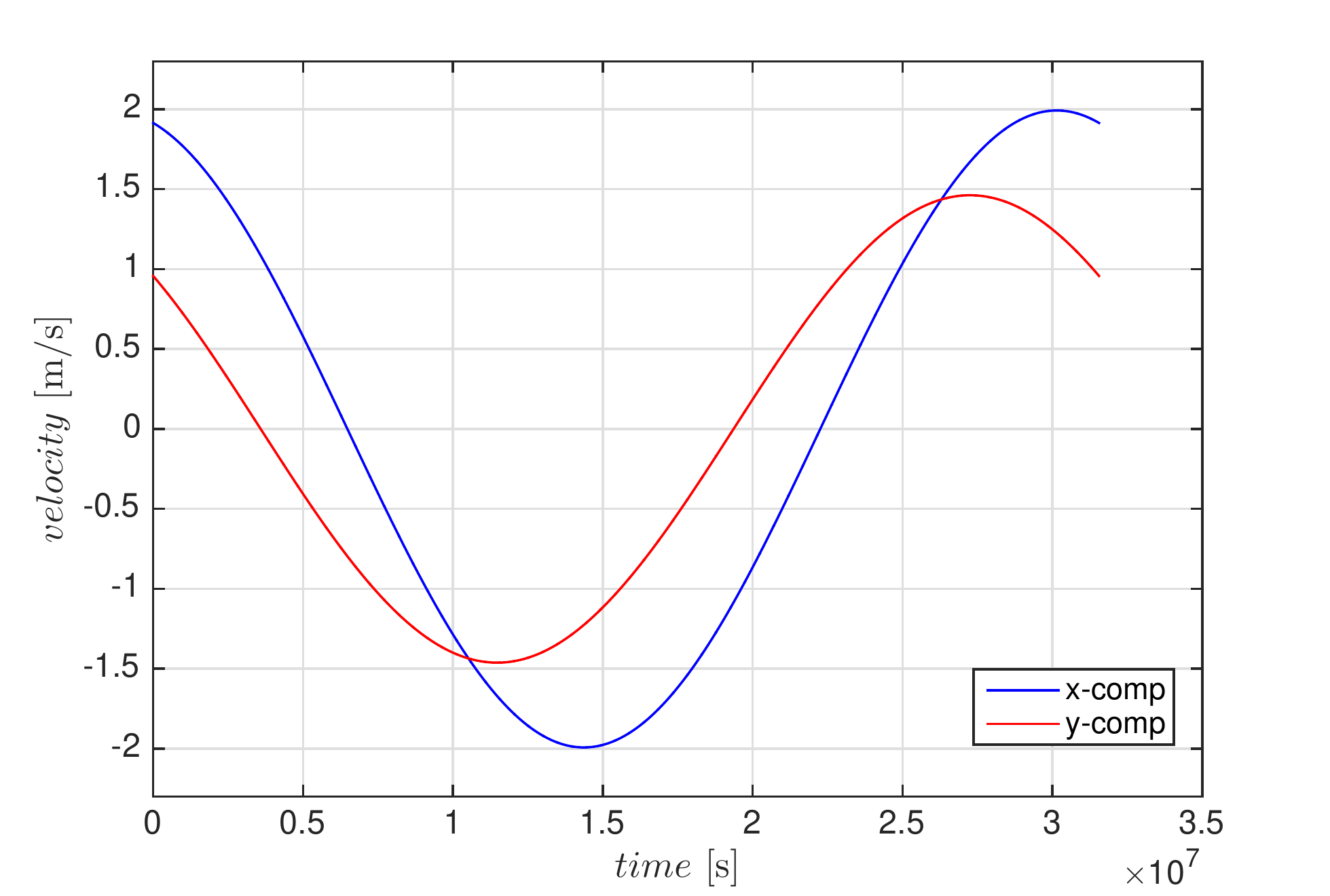}\\
		\caption{Velocity components along the synodic axes for a 10-km-wide orbit around $L_4$.}
		\label{Fig. due}
\end{center}
\end{figure}

Also for the motion around a triangular point, initial conditions can be chosen so as to excite one eigen-frequency, e.g., the fast one (see Fig. \ref{Fig. due} showing the velocity components of a 10-km wide orbit at $L_4$).  In the case of the Sun-Earth system and within the linear approximation, the resulting in-plane and out-of-plane frequencies are practically equal (i.e., with a period of one year) thereby producing a quasi-periodic orbit.

The orbital motion around the Lagrange points will cause a change of the length of the radio-wave path that will overlap with the change associated with the chirality typical of the Lense-Thirring effect.  The question is how to discern one from the other.  The flight time of the radio-waves to cover the $L_2-L_4-L_5-L_2$ circuit is about $2000$ s. During that time the position change of a realistically-sized orbit around a Lagrangian point is greater than the change of the radio-wave circuit path associated with the Lense-Thirring effect.  If the motion around the Lagrangian points were to be purely periodic and with a period that is a fraction of the total duration of the signal data taking (i.e., in order to cover a number of orbital cycles), then this Keplerian-type motion could be resolved from the measured data by frequency analysis. The question that remains to be addressed is how to remove the quasi-periodic, gravity-related components associated with the motion of the spacecraft around the Lagrangian points from the signal, in other words, how to distinguish the (gravity-related) relativistic signal from the non-relativistic secular drift.

Both motions on the stable manifold of the collinear points and around the triangular points can be reconstructed with good accuracy by properly taking into account non-linear effects \cite{Richardson1980,Celletti2015,Paez2015}. Analytic series expansions are obtained in the case of the spatial circular RTBP so to include also out-of-plane motion. Semi-analytical solutions can be further implemented when including more general features like the eccentricity of the primaries \cite{Paez2016}. In both approaches, the quality of the prediction of the time evolution of small amplitude orbits is determined by the order $N$ of the perturbation expansions. The relative error is given by the $N$th power of a perturbative parameter proportional to the amplitude.

One avenue worth exploring for removing the "Keplerian drift" may also hinge on the different behaviors of the relativistic and non-relativistic secular or quasi-secular terms: the Lense-Thirring term grows steadily with time while the Keplerian drift manifests itself as a growth of the orbital amplitude about the Lagrangian points, e.g., either the secular drift of the orbit around $L_1$ and $L_2$ (the orbit slowly spiraling out) or in the case of a triangular point a residual component of the low-frequency term for the orbit around $L_4$ and $L_5$.

In addition to the above points, one should also consider the tracking accuracy in order to separate the relativistic secular signature from the classical effects on the orbit of a spacecraft over the measurement time. The higher possible accuracy is obtained by integrating the Doppler measurements over short arcs. The figure of merit of Doppler measurements in the time domain is well described by means of the Allan deviation $\sigma_y$ \cite{Allan1988}. For instance, considering an Allan deviation of about $3\times 10^{-15}$ (with a reasonable integration time of about $1000$ s), already reached in the case of the tracking of the CASSINI spacecraft \cite{Abbate2002}, the accuracy in the range-rate measurements is about $c\times \sigma_y \sim 9\times10^{-4}$ mm/s, that corresponds to an error in the position of a spacecraft of about $1.8$ mm on the $\sim 2000$ s time span of a single measurement. This error is about $18$ times larger than the precision required on the knowledge of the length of the radiowave circuit (i.e., on the relative position of the satellites) to be compatible with a differential time measurement of $4.3\times10^{-13}$ s. However, by considering $n$ of such short arcs, while both the Lense-Thirring effect to be measured and the knowledge of the relative position between the satellites grow by the same factor $n$, the overall orbit determination accuracy remains at the $1.8$ mm level, making the measurement possible over a time span of about $7.5$ days. By improving the Allan deviation by a factor of three, i.e. $\sigma_y \sim 1\times10^{-15}$, which is possible by current technology, the measurement of the Lense-Thirring effect can be obtained on an overall time span less than $1$ day.

\section{Conclusion}

We have illustrated the proposal of using the system of the Lagrange points of the Sun-Earth system for various experiments and applications. The possibility to measure relativistic time delays both from the Sun and from the Earth has been discussed and the worked out numerical values show that the measurements would be within the range of possibilities offered by current technologies; the experiment would also lend the opportunity to determine the size of the contribution of the quadrupole moment, $J_2$, both of the Sun and of the Earth.

Another proposal we have put forth is the measurement of the inertial frame dragging (Lense-Thirring effect) caused by the angular momentum of the Sun. The technique to be exploited is molded on the Sagnac effect, determining the time of flight asymmetry along a closed path whose edges are the $L$ points, travelled in opposite directions by electromagnetic signals. We have seen that using, for instance, $L_2$, $L_4$ and $L_5$, the time of flight difference would be in the order of a few $10^{-13}$ s, again within the feasibility range of existing technologies. The direct detection of the LT effect of the Sun, besides adding a measurement of a gravito-magnetic phenomenon \textit{per se} to the experiments made in circumterrestrial, or planned in terrestrial, environments, would give the possibility to extract interesting information on the interior of the Sun. We have also discussed the relevance of a possible detection of a galactic gravito-magnetic field; its presence could be evidenced by the envisioned $L$-points configuration at the scale of an AU. A special interest of a possible galactic LT effect is connected with the dark matter halo of the Milky Way, its consistency and, possibly, angular momentum.

Passing to a practical application of the $L$-points set, we have presented and commented a relativistic positioning system at the scale of the full orbit of the Earth. Once more the configuration of the system, its stability in time and its being tied to the orbital motion of the Earth, lend the opportunity of building a positioning and navigation system that could profitably be used by all future space missions, at least in the inner solar system.

Of course all the above is possible provided one can know the actual position of each spacecraft with respect to its $L$-point and keep track of it in time. We have also discussed this fundamental issue and we have seen that a measurement strategy based on continuous data acquisition during few days runs would permit to extract the information on the time of flight asymmetry with the required accuracy.

Once LAGRANGE would have been deployed, there would then indeed be many more opportunities it could offer for fundamental physics, depending on the equipment one would be able to put on board the spacecraft. It is just the case to mention the possibility to detect gravitational waves (GW). The size of the experimental setup would indeed be adequate. An option would be to exploit signals exchanged between the $L$-points adopting a zero-area Sagnac interferometer \cite{Sun1996}. Furthermore, considering again the size and adopting this time a wide area configuration of the light paths (as the triangle $L_2-L_4-L_5$), we should also remember that GW's do carry angular momentum also. The response of the system would strongly depend on the relative orientation, but in principle a GW impinging orthogonally on the ecliptic plane, should superpose a transient asymmetry of the times of flight on the continuous signal due to the solar (and galactic) LT-drag.

Summing up, the idea of using a set of the Lagrangian points of the Sun-Earth system (from two, to four at a time) and measuring the flight times of electromagnetic signals exchanged between spacecraft located in the $L$-points, turns out to be in the range of existing technologies and is appealing. It would be very fruitful for fundamental physics experiments related to tests of GR and possible deviations from it, giving also information concerning the Sun, the Earth and the Milky Way. To pass from proposal to reality an undoubtedly huge effort is required to set up the missions needed to carry and locate the spacecraft at the $L$-points (which could be done progressively, performing different experiments gradually while the stations are launched); to properly equip them; then to control the system and perform the measurements. We think it could be rewarding to try.



\end{document}